# Tumour Induced Angiogenesis and Its Simulation

Sounak Sadhukhan[1*], S. K. Basu[2]

[1,2]Department of Computer Science, Banaras Hindu University, Varanasi 221005, India

**Abstract**

Due to over-metabolism, the tumour cells become hypoxic. To overcome this situation tumour cells secret several chemical substrates to attract nearby blood vessels towards it (angiogenesis). Transition from avascular to vascular tumour is possible with the initiation of angiogenesis. Angiogenesis also plays a crucial role to spread the cancer cells and its colonization at the distant locations of the body (metastasis). In this paper, we briefly review the processes and factors which directly affect tumour angiogenesis or may get affected by it. A model based on cellular automata is developed to demonstrate this complex process through MATLAB based simulation.

## 1. Introduction

New blood vessels formation from the pre-existing blood vessels or by recruiting endothelial precursor cells (EPCs) from the bone marrow is known as neovascularization/angiogenesis. Neovascularization can be characterized into vasculogenic mimicry, and vessel co-option[1,2]. It is a fundamental process that can be seen in some of the physiological processes like reproduction, embryonic development, wound healing, and in some of the pathological processes like, vascularization of ischemic tissues[3], rheumatoid arthritis[4], diabetic retinopathy[5], chronic inflammation, and tumour growth, its progression[6] and metastasis[7], etc.

At the early stage, a tumour nodule does not have any direct vascular support; so it has to depend on the host's vasculature for the nutrients, oxygen, and other growth elements. The metabolic demand of a tumour rises with its volume proportionally[8], but it takes nutrients and oxygen proportionate to its surface area. As a result, cells near the tumour center become hypoxic due to the lack of oxygen and nutrients. At this situation, growth of the tumour would be stagnant (1-2 mm in diameter)[9,10,11]. Without direct access of the blood vessels, the hypoxic tumour cells gradually become necrotic. To continue its growth, the tumour shifts to an angiogenic phenotype and attracts surrounding blood vessels toward it.

During angiogenesis, tumour cells secret a number of chemical species called tumour angiogenic factors (TAFs). These factors diffuse into the microenvironment through extracellular matrix (ECM) and damage the basement membrane of surrounding blood vessels. The TAFs disrupt the corresponding endothelial cell (EC) receptors which directly influence the structure of capillary networks. According to Sholley et al.[12], without proliferation of ECs, capillary network cannot reach the tumour. Therefore, to complete vascularization successfully, ECs must undergo mutation phase. Finger like capillary sprouts are created from the disrupted areas of the surrounding blood vessels[10]. These capillary sprouts grow through the proliferation of recruited ECs from the parent vessels. The TAFs generate a chemical gradient (chemotaxis) between the tumour and the surrounding blood vessels; the capillary sprouts migrate toward the tumour. During angiogenesis, EPCs are brought into the walls of the growing capillary vessels through vascular channel from the bone marrow[13] with the help of angiogenic factors[14], though the percentage of EPCs are very low in general (Fig. 1).

ECs of the capillary sprouts synthesize and secrete cellular fibronectin. This fibronectin does not diffuse in the microenvironment[15] but increases the adhesiveness between the capillary sprouts and the ECM (haptotaxis)[16]. After covering certain distance towards the tumour, tube shaped capillary sprouts often are inclined toward each other and fuse together to form loops (anastomoses). These loops are the milestone of blood circulation within capillary network. From these loops new sprouts are generated and the same sequence of events are repeated to further extend the network. New capillary sprouts can also be generated from a sprout tip (branching). It has been seen that, as the capillary sprouts move toward the tumour, its branching ability rises till the tumour gets penetrated (vascularization). A vascular network is formed from the EC precursors (angioblasts) or from the dual hemopoietic/EC precursors (hemangioblats) in this way. This process is regulated by several pro–angiogenic factors such as, vascular endothelial growth factor (VEGF), acidic- and basic-fibroblast growth factor (a–, b–FGF), placental growth factor (PLGF), angiopoietin–1, 2, and 4 (ANG–1, 2 and 4), etc.[10] and some of the anti–angiogenic factors like angiostatin, endostatin, thorombospondins (TSPs), etc.[17,18]

In physiological angiogenesis, pro- and anti- factors are in balance that regulates the growth of newly developed vessels, whereas, in tumour angiogenesis, the balance is disturbed and pro-factors become more powerful. As a result, newly developed vessels grow unboundedly and the vascular network becomes unorganized, often leaky, chaotic in nature, and haemorrhagic due to upregulation of growth factors. These vessels often carry tumour cells in their wall[19]. The blood flow within these networks is sometimes slow or sometimes oscillating that makes it dysfunctional[20]. Besides sprouting angiogenesis, researchers have identified other mechanisms of angiogenesis in tumours including intussusceptive angiogenesis, vasculogenic mimicry, and lymph–angiogenesis[11], etc.

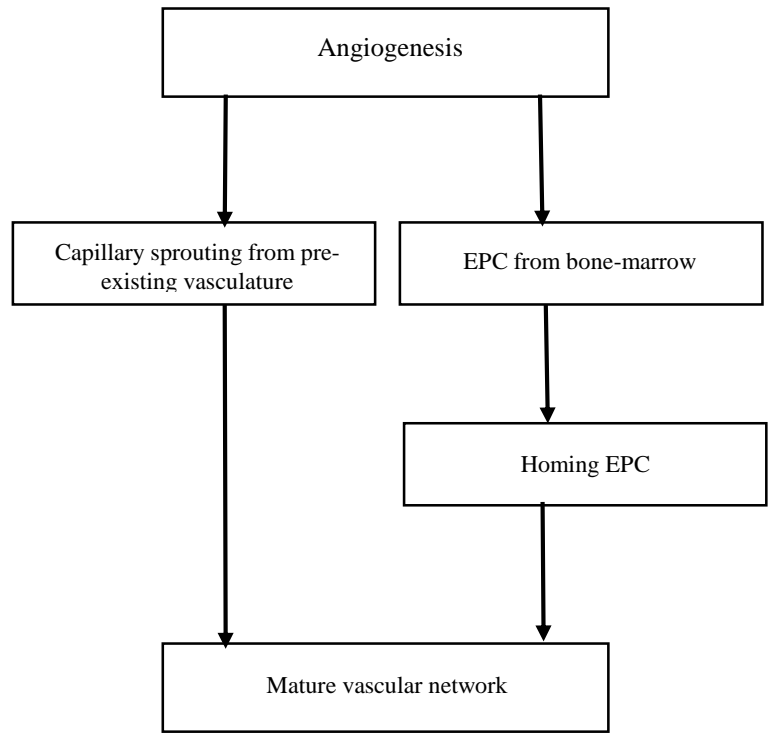

**Fig. 1** Angiogenesis.

In this paper, we review the crucial processes and factors which directly affect tumour angiogenesis (sprouting angiogenesis) or may get affected by it. Moreover, we study tumour angiogenesis from computational oncology perspective. In this context, initially we focus on hypoxia (formed during avascular phase) and its influence on angiogenesis. Mainly the roles of HIF–1α, HIF–2α for influencing TAFs are described. Attention has been paid on the roles of the pro- as well as anti- angiogenic factors like VEGF, FGF, PDGF, ANG, PLGF, TSPs, etc. We also describe the influence of tumour angiogenesis in metastatic process and finally a simple cellular automata (CA) model based on a couple of assumptions/limitations has been described to explain sprouting angiogenesis.

The paper is organized as follows: section 2 briefly describes the effects of hypoxia in angiogenesis; section 3 illustrates the role of TAFs in angiogenesis which include pro- as well as anti- factors; section 4 illustrates the role of angiogenesis in metastatic process; section 5 describes the role of computational and mathematical models for tumour angiogenesis process through a simple CA based modelling; and section 6 concludes the paper.

## 2. Roles of Hypoxia in Tumour Angiogenesis

At the avascular stage, tumour cells require adequate blood supply; otherwise it gradually transforms into hypoxic due to high metabolic consumption and limited supply of oxygen and nutrients. At the hypoxic stage, local pH balance is disturbed, and oxygen and nutrients concentration in the surrounding microenvironment is sharply decreased[21]. The increasing distances between nearby vasculature and the tumour cells due to the tumour mass expansion are also responsible for the hypoxia.

Hypoxia Inducible Factor–1 (HIF–1) is the main controller of the hypoxic response. It belongs to the basic Helix–Loop–Helix PER–ARNT–SIM (b–HLH–PAS) family which consists of αβ–heterodimers. HIF protein family also includes HIF–1α, HIF–2α, HIF–3α, and HIF–1β. Activation of HIF directly affects tumour growth, as it is one of the crucial controllers of blood vessels development, maturation, and remodeling.

Among the three α–subunits of the HIF family, only HIF–1α controls oxygen homeostasis in many cell types[22], and it is a very influencing promoter for tumour growth as it controls EC activity (Fig. 2)[23]. The activity of HIF–1α is controlled by hydroxylation, acetylation, and phosphorylation[24]. It is responsible for transcription of lots of genes which are promoters in angiogenesis (including VEGF), cell proliferation, glucose and iron metabolism, etc.[24] In hypoxic condition, the stability, subcellular localization, and transcriptional activity of HIF–1α are affected. The HIF–1α degradation is prohibited in hypoxia and its levels are accumulated to associate with HIF–1β to use transcriptional roles. In the post–translational modification of HIF–1α, enzymes prolyl–hydroxylase (PHD), and HIF prolyl–hydroxylase (HPH) play an important role and also help HIF–1α to associate with von Hippel–Lindau (VHL) tumour suppressor protein[25]. The activity of PHD depends on the oxygen levels which are also essential for transferring substrates to the proline residue of HIF–1α.

Researchers have found a correlation between HIF–1α and HIF–2α with the activation of many pro–angiogenic factors like VEGF, platelet–derived growth factor–B (PDGF–B), plasminogen activator inhibitor–1 (PAI–1), the Tie–2 receptor, matrix metalloproteinases (MMP–2 and MMP–9), and angiopoiteins (ANG–1 and ANG–2)[26]. These influences destroy the balance between the pro– and anti–angiogenic factors[27]. Among these pro–angiogenic factors VEGF–A is highly expressed in most of the mammalian tumour in ECs[28]. In tumour, elevated VEGF are present at the boundary of necrotic and hypoxic cells. VEGF is responsible for the vessels formation and it has been discovered that HIF–1α acts as a primary regulator for VEGF. The other pro–angiogenic proteins are also regulated by HIF for the vascular development during tumour angiogenesis. For example, HIF activates ANG–1 in ECs[29], which help newly developed matured blood vessels during tumour angiogenesis by recruiting pericytes[30]. HIF–2α targets ANG–2 in ECs, whereas HIF–1α indirectly stimulates ANG–2 expression via VEGF[29]. On the other hand, ANG–1 and ANG–2 both influence Tie–2 receptor, which is a key player in vascular remodelling through destabilizing the existing vessels[31]. Scientists have also discovered that, loss of HIF–1α restricts angiogenesis in ECs including its proliferation and chemotactic migration[23]. Not only in tumour angiogenesis, scientists also got proof that HIF–1α, HIF–2α is also involved in metastasis of different cancers including brain, breast, colon, liver, lung, and etc.[32]

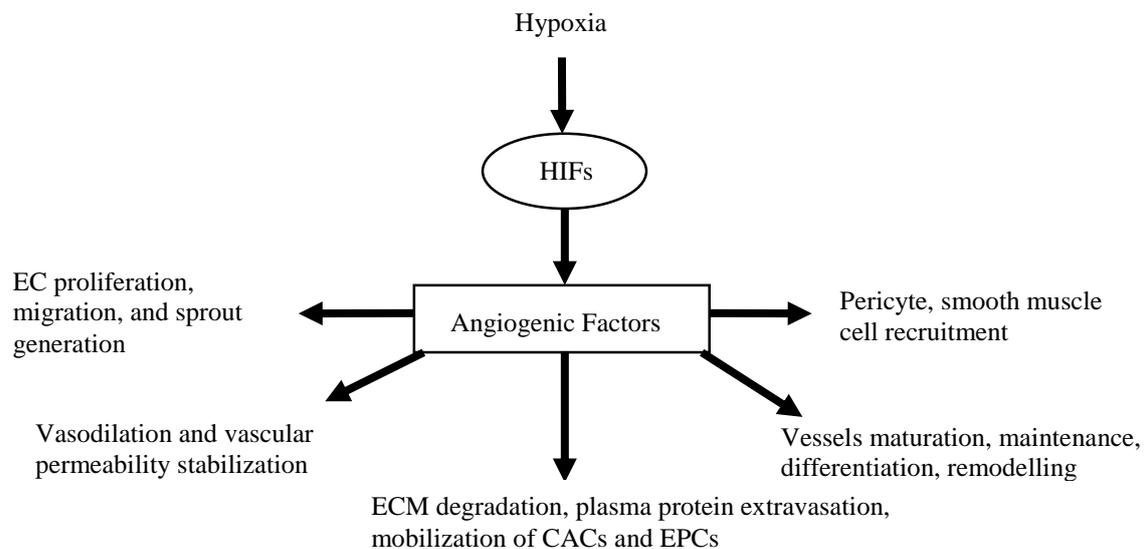

**Fig. 2** HIF–induced angiogenic factors control angiogenic genes, angiogenic phases and functions during tumour angiogenesis (adapted from Krock et al.[33])

## 3. Angiogenetic Factors and Its Influences on Tumour Angiogenesis

In 1968, it was first discovered that the tumour cells secret a number of chemical substrates[34]. In 1971, it was proposed that angiogenesis is an important phase for tumour growth and metastasis[35]. It is also found that prevention in angiogenesis may be a path to restrict the tumour growth. These possibilities motivated the search for angiogenic factors (pro– as well as anti–). Later, it is found that if pro– and anti– factors are in balance, then *angiogenic switch* (Fig. 3) is in *off* state and whenever the balance is lost it is in *on* state triggering angiogenesis in tumour[7].

## 3.1 Pro–angiogenic Factors

There are several factors like metabolic and mechanical stresses, immune response, genetic mutations, etc. which can trigger the *angiogenic switch* into *on* state[36]. The influences of environmental and genetic disorder on tumour angiogenesis and its growth are still unknown. Pro– factors like, VEGF, PLGF, FGF, ANG–1, ANG–2, PDGF, MMPs which are involved in tumour angiogenesis are described below (summarized in Table 1). We discuss about some anti-angiogenic factors in this context also.

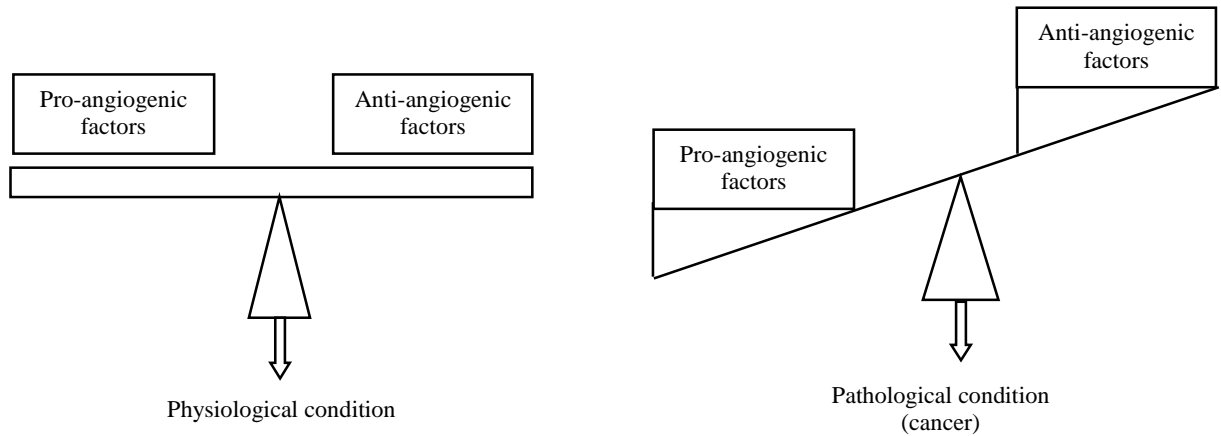

**Fig. 3** The angiogenic switch.

**VEGF/VEGFR**

VEGF, a pro–angiogenic factor, plays important roles in tumour angiogenesis. There are six members in the VEGF family which includes VEGF–A to E, and PLGF that communicate with tyrosine kinases, VEGFR–1, and VEGFR–2 to alter angiogenesis. Scientists have confirmed that the interplay between VEGF–A and VEGFR–2 are the major contributors in sprouting angiogenesis[11]. VEGF–A is a highly specific mitogen for vascular ECs leading to the stimulation of the entire process required for neovascularization. So, it is also named as the vascular permeability factor (VPF). Members of VEGF family bind with tyrosine kinase receptors: VEGFRs, which is activated through trans-phosphorylation. In tumour angiogenesis, bindings between VEGF–A with VEGFR–2 is very crucial. The activities of VEGFR–1 are not well understood yet, though it is known to act as a dummy receptor to isolate VEGF from VEGFR–2 binding. VEGFR–2 is involved in almost all the known cellular responses to VEGF. PLGF is also a promoter of angiogenesis but its activities are less recognized.

It is found that in tumours the level of VEGF is very high. Tumour cells secret VEGF to its surrounding environment. It is used for nourishment of the new blood vessels. At the primary phase of angiogenesis, vascular basement membrane is degraded by VEGF. Then, it binds with the receptors present on the EC surface and activates the proteins like, protein kinase B (PKB/Akt), phosphoinositide 3–kinase (PI–3K), and mitogen–activated protein kinases (MAPK/ERK) which activate EC proliferation. VEGF also up-regulates plasminogen activators[37], and MMPs[38] during tumour angiogenesis. ECM degradation helps EC (of capillary sprouts) migration followed by the cell division to create tube shaped vessels. These newly developed vessels required stability or maturation. Angiopeotin family assures this stability and governs vascular growth.

In the case of normal vessels, proper involvement of pericyte decreases EC proliferation and also reduces their dependence on the host tissue production of VEGF-A. On the contrary, improper association of pericyte may cause abnormal vessel diameter and sensitivity to VEGF inhibition, which may lead to apoptosis of ECs. Recent studies claim that ANG-1 and PLGF can provide survival signals and they have the ability to rescuing immature blood vessels from VEGF-A loss.

**FGF/FGFR**

FGFs are heparin binding factors. It binds with receptor tyrosine kinases (RTKs). FGFs and its receptors (FGFRs) control growth and differentiation of normal cells. FGF family consists of 23 members. Among them, a–FGF (acidic) (FGF–1), and b–FGF (basic) (FGF–2) are identified as the pro–angiogenic factors which influence EC proliferation and migration[39]. The up-regulation of FGF and FGFR are found in tumour cells. It is a very important factor for the angiogenesis cascade, though it does not have signaled sequences that trigger secretion. FGF is secreted in ECM; after that angiogenesis starts. In angiogenesis, FGF works with VEGF in co–operated manner. So, the interaction between FGF and VEGF is also an important issue in tumour angiogenesis[40].

**ANG-1, 2, 4 /Tie–2**
The angiopoietin family consists of four members (identified till now), among which ANG–1, ANG–2, and ANG–4 are involved in angiogenesis. ANG–1, 2, and 4 bind with endothelial tyrosine kinase receptor: Tie–2. Angiopoietin family restricts several ligands to bind with the same receptors. While ANG–1 activates Tie–2 signal, ANG–2 opposes its activation. The functions of ANG–1 are EC migration, adhesion, and pericyte recruitment; on the other hand ANG–2 destabilizes the vessels[41].

**Table 1** Essential pro-angiogenic factors, their receptors, and functions

| Angiogenic Factors | Receptors | Functions |
|---|---|---|
| VEGF | VEGFRs (VEGFR–1, VEGFR–2) | *initiation of angiogenesis*<br>1. permeability of vessels<br>2. pericytes detachments<br>3. basement membrane degradation<br>*neovascularization*<br>4. EC proliferation and migration<br>5. pericyte proliferation and migration |
| PLGF | VEGFR–1 | *neo–vessels formation*<br>1. proliferation and migration of ECs |
| FGF | FGFRs | *neo–vessels formation*<br>1. proliferation and migration of ECs<br>2. proliferation and migration of pericytes<br>3. basement membrane degradation |
| ANG–1 | Tie–2 | *maturation*<br>1. accumulate EC and lumen<br>2. pericytes attachment<br>3. basement membrane deposition<br>4. maintenance of vessels |
| ANG–2 | Tie–2 | *initiation of angiogenesis*<br>1. pericytes detachments<br>2. basement membrane degradation<br>*adaptation to tissue needs*<br>1. regression of neo vessels due to lack of flow or presence of growth factor |
| PDGF–B | PDGFR | *neo–vessels formation*<br>1. proliferation and migration of ECs<br>2. proliferation and migration of pericytes<br>*maturation*<br>1. pericytes attachment<br>2. basement membrane deposition |

**PDGF/PDGFR**
The PDGF family consists of five members: PDGF–A to D, and PDGF–AB. These proteins activate the downstream signals through binding PDGFR tyrosine kinases receptors and activating molecules like VEGF[42,43]. PDGFs are mainly involved in growth and progression of various kinds of cells. Scientists found a strong correlation between the up-regulation of PDGF, with FGF and VEGF[44,45]. PDGFs are involved in the recruitment of pericytes to the blood vessels, enhancement of secretion of other pro–angiogenic factors, endothelial proliferation and motility, sprouting and tube formation, promotion of lymph-angiogenesis, and subsequent lymphatic metastasis[46].

Besides the described factors, cancer cells produce other factors including transformation growth factor–β (TGF–β), epidermal growth factor (EGF), etc. TGF–β acts as a stabilizer of newly developed blood vessels and suppressor of the immune system[47]; and EGF up-regulates VEGF[48]. Other major factors for angiogenesis are MMPs. It helps to degrade the basement membrane of vessel's wall and remodels ECM[49].

## 3.2 Anti–angiogenic Factors
In tumour angiogenesis, pro-aniogenic factors are more up-regulated than anti-angiogenic factors. Anti-angiogenic factors also exist in this mechanism to inhibit angiogenesis. Some of the anti-angiogenic factors are angiostatin, endostatin, thorombospondins (TSPs) etc.[17,18]

**Angiostatin**

It consists of one or more plasminogen fragments. This plasminogen is formed by proteolytic cleavage. Researchers showed that the combination of angiostatin and endostatin possess anti-tumour property against glioblastoma cells *in vitro* as well as *in vivo* and it increases proliferation, decreases micro-vessel density, enhances survival time of tumour and reduces tumour volume[50].

**Endostatin**
It is a naturally developed anti-angiogenic factor from proteolytic cleavage (from XVIII collagen). It may communicate with multiple cell-surface molecules. Scientists have shown that the combination of endostatin with chemotherapy can enhance progression free survival and also improve clinical response in the case of advanced sarcomas patients[51].

**TSPs**
TSPs are found to inhibit angiogenesis which delays tumour progression. Among the TSPs family TSP-1 is very important in tumour angiogenesis. Some studies have said TSP-1 has positive correlation with p53 in several types of tumours as p53 genes up-regulate TSP-1, down-regulate angiogenesis and suppress the tumour[52,53]. The roles of p53 in tumour angiogenesis are summarized in Table 2.

Table 2 Roles of p53 genes in angiogenesis[20]

| Factors up-regulated by p53 | |
|---|---|
| TSP-1 | Endogenous angiogenesis inhibitors |
| MMP-2 | Promotes endogenous anti-angiogenic peptide formation from basement-membrane proteins |
| Brain specific angiogenesis inhibitors | Brain specific endogenous angiogenesis inhibitors |
| Eph receptor A2 (EphA2) | EphA2 induces apoptosis; soluble EphA2 receptors are anti-angiogenic |
| Factors down-regulated by p53 | |
| MMP-1 | Promotes angiogenesis in bone and activates latent MMP-2 |
| VEGF-A | Potent pro-angiogenic and permeability factor |
| HIF–1α | Induces angiogenesis by activation of VEGF-A transcription under hypoxic conditions |
| Cyclooxygenase-2 | Promotes angiogenesis |

## 4. Roles of Angiogenesis in Cancer Spread

Angiogenesis is an essential phase for the tumour cell to metastasize. There is a very strong correlation among tumour micro vessels density (MVD), metastatic potential, and the poor survival rates[54,55,56]. In tumour angiogenesis, newly developed capillaries grow unboundedly. As a result, the vascular network becomes unorganized, often chaotic in nature, leaky and haemorrhagic due to up-regulation of growth factors. These vessels often carry tumour cells in their wall[19]. The leaky vessels enable hematogenous spread of cancer cells either singly or collectively[57]. After detaching from the primary tumour, cancer cells may enter into the leaky vessels and spread all-over the body to metastasize at remote sites.

Due to angiogenesis, tumour gets direct access to the blood vessels, and absorbs oxygen and nutrients as much as possible for its unbounded proliferation. On the other hand, tumour secrets matrix degrading enzymes like, MMPs which help to remodel ECM[49]. Damaged ECM secrets a number of ECM-associated growth factors which also help tumour cells to grow faster and invade into the other tissues. As a result, tumour grows in volume and produces tremendous biomechanical pressure on the adjacent tissues including basement membrane, which gets dislocated and compressed. This phenomenon creates extra spaces for the tumour to grow and form "finger" like protrusions of tumour cells that invade into the surrounding tissue along the minimum resistant path through ECM. The ability to invade surrounding tissues differentiates a malignant tumour from a benign tumour.

Angiogenesis is not only providing nutrients and oxygen for the tumour cells but also it provides tumour cells a route to enter into the blood vessels or lymphatic vessels (intravasation). In intravasation, tumour cells detach from its neighbourhood and invade into the capillary vessels. It penetrates the basement membrane of vessels first and then damages EC barrier to enter into the blood stream with the help of transitional mesenchymal phenotype. These routes the tumour cells into the circulatory system. After invading into the blood and/or the lymphatic vessels, tumour cells collectively or singly spread to the other locations of the body through circulation in vessels. During the migration through the circulatory system, a lot of forces like, haemodynamic force, immunological stress, collisions with blood cells and ECs (of vessel wall) act on tumour cells together. Research showed that after 24 hours in circulation, less than 1 percent of cancerous cells are viable and less than

0.1 percent survived to form metastasis[58]. The circulating tumour cells which succeed to survive from fluid shear and immune system of the body may get arrested to the vascular endothelium of secondary site and exit from the circulatory system to go into the tissue (extravasation). But, extravasation does not necessarily mean that the tumour cell is metastasizing. Singly migrated tumour cells that are extravasated mostly either apoptose, killed by immune system, or dormant[59]. After extravasation, tumour cell often make a colony of cells at the secondary site. This phenomenon is called metastasis. The secondary tumour also proliferates and undergoes angiogenesis to become fatal (macro-metastases). This metastatic tumour may re-metastasize to another site or can even re-implant in the primary tumour[60]. Almost ninety percent patients suffering from cancer have died of metastasis.

## 5. Tumour Angiogenesis: Computer Modelling Perspective

Computational oncology involves computer-based modelling relating to tumour biology and cancer therapy. The aim is to build computer models that simulate biological processes and to use these models to make useful predictions. Scientists have developed various models to explain tumour growths as well as its progression from different perspectives as most of the processes involved in tumour growth are not clearly understood. A few computational and mathematical models of tumour angiogenesis are briefly discussed below. We also describe our own CA based discrete model for tumour angiogenesis.

The work of Anderson and Chaplain[61] is one of the most cited research paper in tumour angiogenesis modeling. In that study, they have presented both continuum as well as discrete models for capillary sprout formation in tumour angiogenesis. Both of these models consider the interactions between cell-ECM via fibronectin. They have concluded that *chemotaxis* as well as *haptotaxis* are the necessary factors for the development of capillary network and the chemotactic response should be strong enough for the initial outgrowth of capillary network. Milde et al.[62] have developed a hybrid model of sprouting angiogenesis. The model combines continuum approximation of several TAFs, fibronectin, and EC density with discrete particle based sprout tip cells. The capillary vessels are represented by the continuous grid-independent particle mesh interpolation method. Also, filopodia are modeled explicitly to sense the migration and branching direction. The model explains that the structure and the density of the ECM has direct impact on the morphology, EC migration, expansion speed, the number of capillary branches, and network formation in the newly developed vessels.

Wcisło et al.[63] have developed a CA based 3-D multi-scale model based on particle dynamics for tumour angiogenesis. The tissue and fragments of vascular network are formed with particles and interact with the surrounding neighbors via mechanical resistance of the cell wall. The model was able to explain the realistic 3-D dynamics of the entire tumour microenvironment consisting of normal and cancerous cells, blood vessels, and blood flow. Owen et al.[64] have also developed a 3-D multi-scale model for tumour induced angiogenesis which includes vascular tissue growth with blood flow, capillary sprout formation, and vascular remodeling. The model has illustrated that the blood vessel pruning varies with the pressure drop across a vascular network proportionally. The model has explained the role of hypoxia in angiogenesis and also showed that the level of oxygen in blood controls the blood vessel generation. Shirinifard et al.[65] have developed a 3-D cellular-pott model to mimic the solid tumour growth and angiogenesis by neglecting a few facts for example, anastomosis; possible presence of pericytes and ECM; dynamism among tumour cells; blood vessels attributes like, vessel diameter, blood flow rate, and vessel collapse due to external pressure; vessels remodeling etc. Lyu et al.[66] have developed a hybrid model for tumour angiogenesis. The model has described the tumour growth process and the changes in the microenvironment from the avascular to the vascular stage. The study has well justified the physiological facts.

In the next sub-section, CA based model for sprouting angiogenesis is proposed in 3-D framework by neglecting the dynamic nature of the tumour cells, dynamics of oxygen and nutrient concentrations, presence of pericytes, presence of anti-angiogenic factors, differences between the veins and arteries, and smooth muscle cells.

### 5.1 CA Model for Tumour Angiogenesis

A CA model is proposed for tumour angiogenesis based on a couple of assumptions. The growth of a tumour becomes stagnant after it becomes 1-2 mm. in diameter[9,10,11]. In our model, it is assumed that the tumour is a static entity by ignoring the changes in tumour cell dynamics. Once the tumour cells secret TAFs, it diffuse through the ECM and remodel the basement membrane of the nearby blood vessels. After stimulation by the TAFs, EC lining of the blood vessels form finger like sprouts which grow by recruiting the ECs of the parent vessels and move toward the tumour due to the higher concentration of TAFs. The capillary sprout tips (*S*) control the movement of these newly created capillaries. ECs secret fibronectin, creating adhesive force between the capillary sprouts and the ECM. On the other hand capillary sprouts secrete MMPs (upregulated by the TAFs), which remodel ECM. This also helps EC sprouts to make their way through the ECM.

In our simulation, we consider that the tissue is represented by a 3-D grid of cells. The domain size of CA is [0, 1] X [0, 1] X [0, 1]. We assume that the spatial domain is discretized as $x = ih$, $y = jh$, $z = kh$ (space step $h$), and the temporal domain $t = n\tau$; where $i, j, k, h, n$, and $\tau$ are positive parameters. We consider that the spatial domain length is 2 mm. and the step size $h = 0.01$. Hence, a unit of step size represents 20 μm of dimensional length which is equivalent to the diameter of four ECs.

We use Moore's as well as von Neumann's neighborhoods. $M(H_{i,j,k})$ refers to the Moore's neighborhood elements of $(i, j, k)^{th}$ position in an arbitrary 3-D grid $H$, that is, $H_{i+1,j,k}$; $H_{i-1,j,k}$; $H_{i,j+1,k}$; $H_{i+1,j+1,k}$; $H_{i-1,j+1,k}$; $H_{i-1,j-1,k}$; $H_{i,j-1,k}$; $H_{i+1,j-1,k}$; $H_{i,j,k+1}$; $H_{i+1,j,k+1}$; $H_{i-1,j,k+1}$; $H_{i,j+1,k+1}$; $H_{i+1,j+1,k+1}$; $H_{i-1,j+1,k+1}$; $H_{i-1,j-1,k+1}$; $H_{i,j-1,k+1}$; $H_{i+1,j-1,k+1}$; $H_{i,j,k-1}$; $H_{i+1,j,k-1}$; $H_{i-1,j,k-1}$; $H_{i,j+1,k-1}$; $H_{i+1,j+1,k-1}$; $H_{i-1,j+1,k-1}$; $H_{i-1,j-1,k-1}$; $H_{i,j-1,k-1}$; $H_{i+1,j-1,k-1}$. Whereas $N(H_{i,j,k})$ refers to the von Neumann neighborhood of $(i, j, k)^{th}$ position in $H$, that is $H_{i+1,j,k}$; $H_{i-1,j,k}$; $H_{i,j+1,k}$; $H_{i,j-1,k}$; $H_{i,j,k+1}$; and $H_{i,j,k-1}$. In this model, we use several TAFs concentrations like VEGF ($V$), FGF ($F$), PDGF ($P$), ANG ($A$), etc., ECs ($e$), fibronectin ($f$), MMPs ($m$), and ECM ($E$). From the above considerations, the following CA rules (1-9) are formulated.

$$V_{i,j,k}^{n+1} = V_{i,j,k}^n + \tau\left(\alpha_1 avg.\left(M(V_{i,j,k}^n)\right) - \alpha_2 V_{i,j,k}^n avg.\left(e_{i,j,k}^n + M(e_{i,j,k}^n)\right)\right) \qquad (1)$$

$$F_{i,j,k}^{n+1} = F_{i,j,k}^n + \tau\left(\alpha_3 avg.\left(M(F_{i,j,k}^n)\right) - \alpha_4 F_{i,j,k}^n avg.\left(e_{i,j,k}^n + M(e_{i,j,k}^n)\right)\right) \qquad (2)$$

$$P_{i,j,k}^{n+1} = P_{i,j,k}^n + \tau\left(\alpha_5 avg.\left(M(P_{i,j,k}^n)\right) - \alpha_6 P_{i,j,k}^n avg.\left(e_{i,j,k}^n + M(e_{i,j,k}^n)\right)\right) \qquad (3)$$

$$A_{i,j,k}^{n+1} = A_{i,j,k}^n + \tau\left(\alpha_7 avg.\left(M(A_{i,j,k}^n)\right) - \alpha_8 A_{i,j,k}^n avg.\left(e_{i,j,k}^n + M(e_{i,j,k}^n)\right)\right) \qquad (4)$$

$$f_{i,j,k}^{n+1} = f_{i,j,k}^n + \tau(\omega_1 e_{i,j,k}^n - \omega_2 f_{i,j,k}^n e_{i,j,k}^n) \qquad (5)$$

$$m_{i,j,k}^{n+1} = m_{i,j,k}^n + \tau\left(\gamma_1 avg.\left(M(m_{i,j,k}^n)\right) + \gamma_2 \phi(i,j,k) e_{i,j,k}^n - \gamma_3 E_{i,j,k}^n m_{i,j,k}^n\right) \qquad (6)$$

where $\phi(i,j,k) = \begin{cases} 1 & \text{If } (i, j, k) \text{ represents a sprout tip at the time instance } n. \\ 0 & \text{Otherwise.} \end{cases}$ (6.1)

$$E_{i,j,k}^{n+1} = E_{i,j,k}^n - \tau \lambda avg.\left(M(m_{i,j,k}^n)\right) E_{i,j,k}^n \qquad (7)$$

$$Direction\ of\ (S_{i,j,k}^{n+1}) = index\ of\ max(\psi_{i,j,k}^n, \psi_{i+1,j,k}^n, \psi_{i-1,j,k}^n, \psi_{i,j+1,k}^n, \psi_{i,j-1,k}^n, \psi_{i,j,k+1}^n, \psi_{i,j,k-1}^n) \qquad (8)$$

where $\psi_{i,j,k}^n = \underbrace{(\mu_1 V_{i,j,k}^n + \mu_2 F_{i,j,k}^n + \mu_3 P_{i,j,k}^n + \mu_4 A_{i,j,k}^n)}_{\text{chemotaxis}} + \underbrace{w_2 e_{i,j,k}^n f_{i,j,k}^n}_{\text{haptotaxis}} + \underbrace{w_3 p_{i,j,k}^n d_{i,j,k}}_{\text{random motility}}$ (8.1)

$$e_{i,j,k}^{n+1} = e_{i,j,k}^n + \tau\left(\beta\left(\chi(i,j,k) \times avg.\left(N(e_{i,j,k}^n)\right)\right)(\mu_1 V_{i,j,k}^n + \mu_2 F_{i,j,k}^n + \mu_3 P_{i,j,k}^n + \mu_4 A_{i,j,k}^n)\right) \qquad (9)$$

where $\chi(i,j,k) = \begin{cases} 1 & \text{If any von Neumann neighbor of } (i, j, k) \text{ belong to an EC at } n \\ 0 & \text{Otherwise.} \end{cases}$ (9.1)

In these Equations (1 – 9), $\alpha_1$ through $\alpha_8$, $\omega_1$, $\omega_2$, $\gamma_1$ through $\gamma_3$, $\lambda$, $w_1$ through $w_3$, $\mu_1$ through $\mu_4$, and $\beta$ are the positive parameters having values between 0 and 1. $avg.\left(M(V_{i,j,k}^n)\right)$ represents the average value of Moore's neighborhood of VEGF concentration, and $avg.\left(N(e_{i,j,k}^n)\right)$ symbolizes the average value of von Neumann's neighborhood of EC concentration at the time instance $n$. In the Equation (8.1), $p_{i,j,k}^n$ is the randomly generated value between 0 to 1 at the time instance $n$, and $d_{i,j,k}$ is the reciprocal of the distance between the tumour center and the point $(i, j, k)$. If Equation (8) returns the position $(i, j, k)$, then it indicates that the sprout tip will be in the idle position otherwise the sprout moves to a neighboring specified position if and only if the position is not pre-occupied.

The rule of anastomosis and branching is not clearly known. So, for this simulation, we consider that anastomosis occurs if and only if two sprout tips are situated at the adjacent positions (von Neumann neighbour) of the grid and $P_a$ is the chances of anastomoses. On the other hand, conditions for branching are as follows,
  i) The sprout tip should be mature ($age \geq age_{th}$).
  ii) There should be sufficient space available for branching.
  iii) The probability for branching of a sprout tip is $P_b$.
Capillary sprout branching can happen if and only if these three conditions are true simultaneously.

## 5.2 Parameters Estimation and Simulation

In this simulation, each and every component is initialized with the proper values. For example,

$$\left.\begin{aligned}
V(x,y,z,0) &= \sigma_1 e^{\frac{-(1-y)^2}{\varepsilon_1}}, (x,y,z) \in [0,1] \times [0,1] \times [0,1] \\
F(x,y,z,0) &= \sigma_2 e^{\frac{-(1-y)^2}{\varepsilon_2}}, (x,y,z) \in [0,1] \times [0,1] \times [0,1] \\
P(x,y,z,0) &= \sigma_3 e^{\frac{-(1-y)^2}{\varepsilon_3}}, (x,y,z) \in [0,1] \times [0,1] \times [0,1] \\
A(x,y,z,0) &= \sigma_4 e^{\frac{-(1-y)^2}{\varepsilon_4}}, (x,y,z) \in [0,1] \times [0,1] \times [0,1] \\
\text{and } f(x,y,z,0) &= \sigma_5 e^{\frac{-y^2}{\varepsilon_5}}, (x,y,z) \in [0,1] \times [0,1] \times [0,1]
\end{aligned}\right\} \quad (10)$$

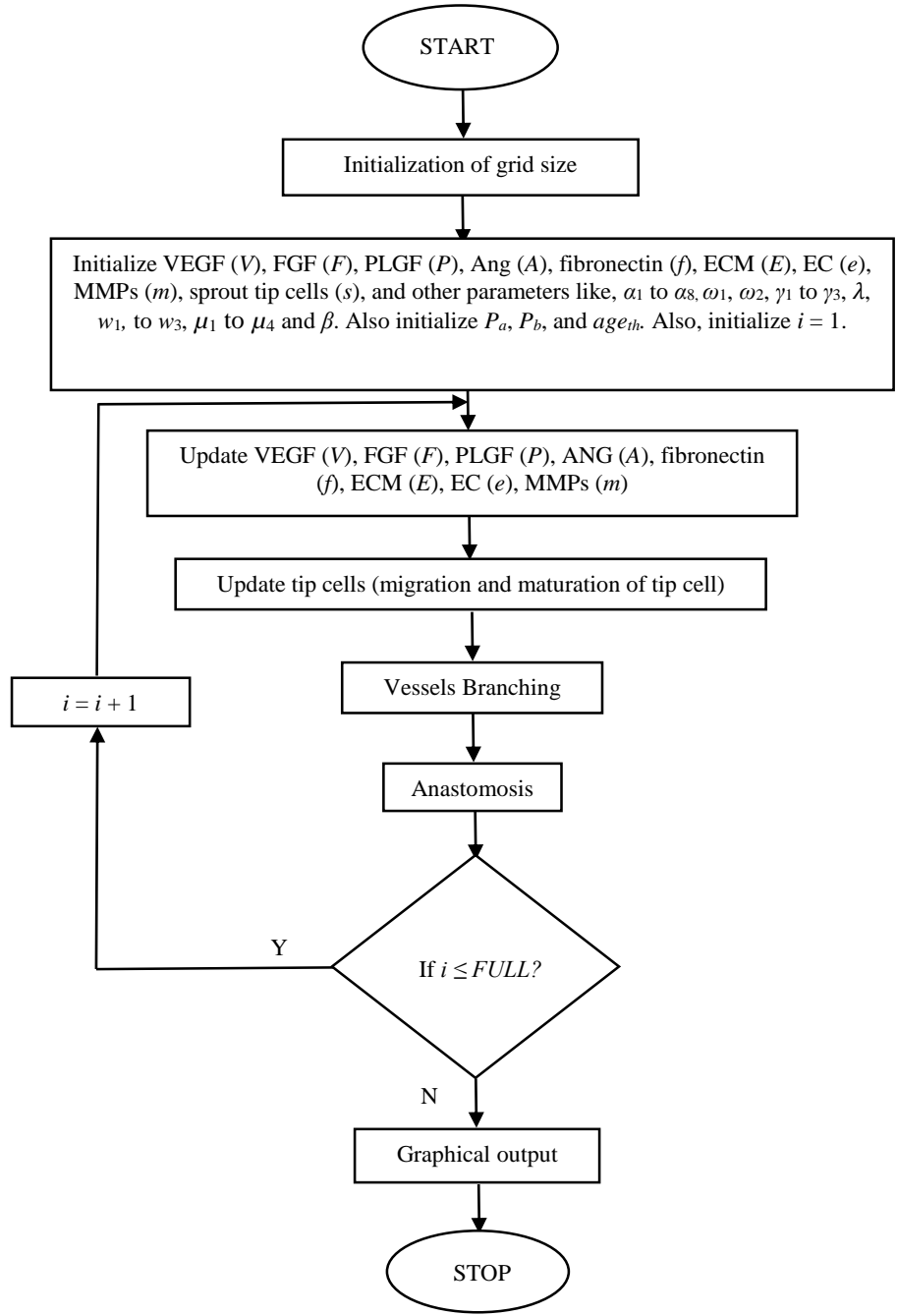

**Fig. 4** Flow chart of CA based simulation for tumour angiogenesis.

where $\sigma_1 = 1.0$, $\sigma_2 = 0.95$, $\sigma_3 = 0.90$, $\sigma_4 = 0.85$, $\sigma_5 = 1.0$, $\varepsilon_1 = 0.45$, $\varepsilon_2 = 0.40$, $\varepsilon_3 = 0.35$, $\varepsilon_4 = 0.30$, and $\varepsilon_5 = 0.45$. The ECM is initialized with random values between 0 and 1. We assume that, the tumor centre is located at (0.5, 0, 0.5) and the radius of the tumour is 1 mm or 50$h$. The pre-existing blood vessels are represented by the horizontal straight lines (0, 0, 0.2) to (1, 0, 0.2); (0, 0, 0.4) to (1, 0, 0.4); (0, 0, 0.6) to (1, 0, 0.6); (0, 0, 0.8) to (1, 0, 0.8); and the vertical straight lines (0.3, 0, 0) to (0.3, 0, 1) and (0.6, 0, 0) to (0.6, 0, 1), respectively. The new capillary sprouts can be generated randomly from any point on these blood vessels. We also consider that initially MMPs are not present in the spatial domain, but the tip cell of capillary sprouts secrets MMPs, hence, at $t = 0$ MMPs is initialized with 0.

The level of concentrations of all the substrates mentioned above are non-dimensional and scaled between 0 and 1. In the updation rules defined by the Equations (1 – 9), the parameter values used for simulations are $\alpha_1 = 0.001$, $\alpha_2 = 0.1$, $\alpha_3 = 0.001$, $\alpha_4 = 0.09$, $\alpha_5 = 0.001$, $\alpha_6 = 0.08$, $\alpha_7 = 0.001$, $\alpha_8 = 0.075$, $\omega_1 = 0.05$, $\omega_2 = 0.1$, $\gamma_1 = 0.01$, $\gamma_2 = 0.009$, $\gamma_3 = 0.0075$, $\lambda = 0.1$, $\beta = 0.1$, and $\tau = 1$ (equivalent to 2.4 Hrs.). We consider that a sprout maturation time ($age_{th}$) is $10\tau$ (equivalent to 24 Hrs.).

All the grid elements are updated concurrently. At each time step, we update the values of VEGF, FGF, PDGF, ANG concentrations, ECs density, fibronectin, MMPs, capillary sprout tips, and the ECM. The algorithm (Fig. 4) is executed with $w_1 = 0.30$, $w_2 = 0.40$, and $w_3 = 0.30$ to observe the effects of *chemotaxis*, *haptotaxis*, and *random motility* on the sprout tip-cell movement on 5th, 10th, 15th, and 20th day (Fig. 5 (a)-(d)) respectively.

Through the simulation described above, a number of observations are made: a) no two capillary sprouts are generated from adjacent locations; b) as the sprouts approach closer to the tumour, its branching tendency increases, c) *chemotaxis* is the most effective driving force than *haptotaxis* as there is no significant changes visible with $w_2 = 0$.

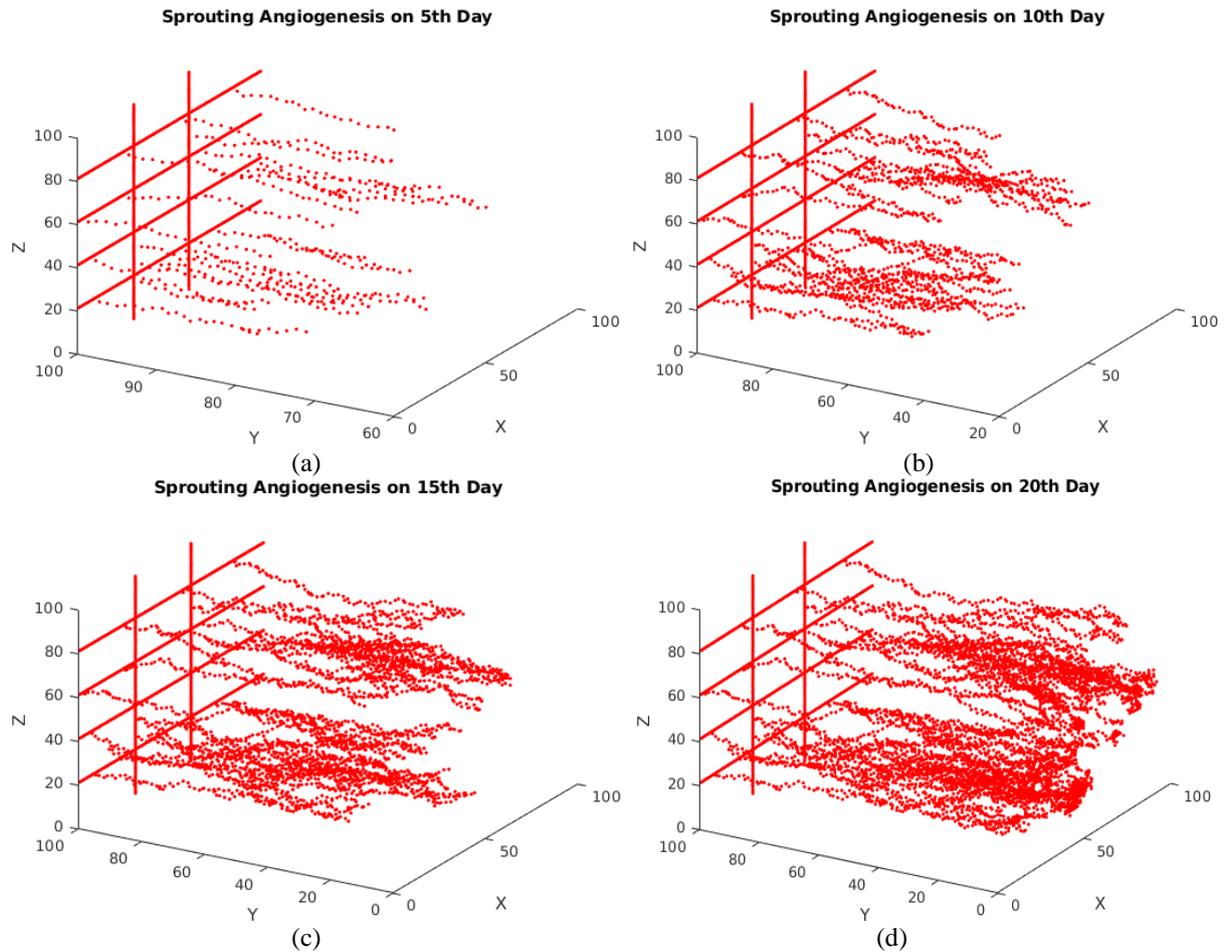

**Fig. 5** Capillary sprout formation in tumour induced angiogenesis due to *chemotaxis*, *haptotaxis*, and *random motility* on (a) 5th, (b) 10th, (c) 15th, and (d) 20th day respectively.

## 6. Conclusions

Angiogenesis is a critical phase in cancer as it acts as a bridge between avascular and vascular stage. To overcome the hypoxic state, tumour cells switch into angiogenic phenotype to attract nearby blood vessels. It is regulated by pro- and anti- factors. Tumour angiogenesis also acts as a medium for metastatic spread of cancer cells at the distant locations. It was hypothesized that if somehow angiogenesis would get inhibited then the aggressiveness and the progression of a tumour would be reduced. This was the motivation for active research on tumor angiogenesis. Unfortunately, very less information is available regarding the processes involved in tumour angiogenesis.

    To overcome this situation, computer-based modelling can be used. The aim is to build a computer model that simulates tumour angiogenesis to make useful predictions. In order to achieve this objective, a CA based simple model is proposed to demonstrate this phenomenon. Using the interactions in 3-D tissue space, we study the effects of various biological factors like, VEGF, FGF, PDGF, ANG, MMPs, fibronectin, and ECM on the capillary sprouts formation. The model also captures anastomoses and branching phenomena in capillary blood vessels. In each run of the simulation, the number of sprout formation and locations are varied randomly. Overall, the model justifies the practical scenario though it needs to be examined further with more factors and more details by including dynamic nature of the tumour cells, dynamics of oxygen and nutrient concentrations, capillary vessels pruning, and blood flow through the newly formed vessels.

## Acknowledgement

The first author of this paper is thankful to the University Grant Commission, Government of India for supporting him with a Senior Research Fellowship.